\documentclass[aps,prl,twocolumn,superscriptaddress]{revtex4}
\usepackage{graphicx}
\usepackage{amsmath,amssymb}
\usepackage{bm}
\usepackage{verbatim}
\usepackage{ulem}

\begin{document}

\title{Strong Interaction Effects in Superfluid Ising Quantum Phase Transition}
\author{Wei Zheng}
\affiliation{Institute for Advanced Study, Tsinghua University, Beijing, 100084, China}
\author{Boyang Liu}
\affiliation{Institute for Advanced Study, Tsinghua University, Beijing, 100084, China}
\author{Jiao Miao}
\affiliation{Institute for Advanced Study, Tsinghua University, Beijing, 100084, China}
\author{Cheng Chin}
\affiliation{James Franck Institute, Enrico Fermi Institute and Department of Physics, University of Chicago, Chicago, Illinois, 60637, USA}
\author{Hui Zhai}
\email{hzhai@mail.tsinghua.edu.cn}
\affiliation{Institute for Advanced Study, Tsinghua University, Beijing, 100084, China}
\date{\today}

\begin{abstract}
We study the quantum phase transition between a normal Bose superfluid to one that breaks additional $Z_2$ Ising symmetry. Using the recent shaken optical lattice experiment as an example, we first show that at mean-field level atomic interaction can significantly shift the critical point. Near the critical point, bosons can condense into a momentum state with high or even locally maximum kinetic energies due to interaction effect. Then, we present a general low-energy effective field theory that treats both the superfluid transition and the Ising transition in a uniform framework, and identify a quantum tricritical point separating normal superfluid, $Z_2$ superfluid and Mott insulator. Using perturbative renormalization group method, we find that the quantum phase transition belongs to a unique universality class that is different from that of a dilute Bose gas.
\end{abstract}

\maketitle

Critical phenomena lie in the center of modern many-body physics. Near the phase transition, the many-body system can develop universal and unconventional behaviors. Two of the most paradigmatic phase transitions are Ising transition and superfluid transition.  Across an Ising transition a discrete $Z_2$ symmetry is spontaneously broken, while a $U(1)$ gauge symmetry is spontaneously broken across a superfluid transition. If a system can exhibit phase transitions of both two types of symmetry breaking, their interplay can lead to novel critical phenomena. Such system is not known in real materials, to the best of our knowledge, but has been recently demonstrated in cold atom systems.

So far there are at least three approaches to realize such a transition in cold atom experiments: a) Bose condensates with spin-orbit coupling induced by Raman transitions, where the transition is driven by changing the Raman coupling strength \cite{NIST,USTC}; b) Bose condensates in an optical lattice with staggered magnetic field, where the transition is driven by changing the ratio of two hopping amplitudes along two different spatial directions \cite{mag2}; and c) Bose condensates in a shaking optical lattice, where the transition is driven by tuning the shaking frequency \cite{Sengstock1} or shaking amplitude \cite{Cheng}.

\begin{figure}[tbp]
\includegraphics[width=3.0in]
{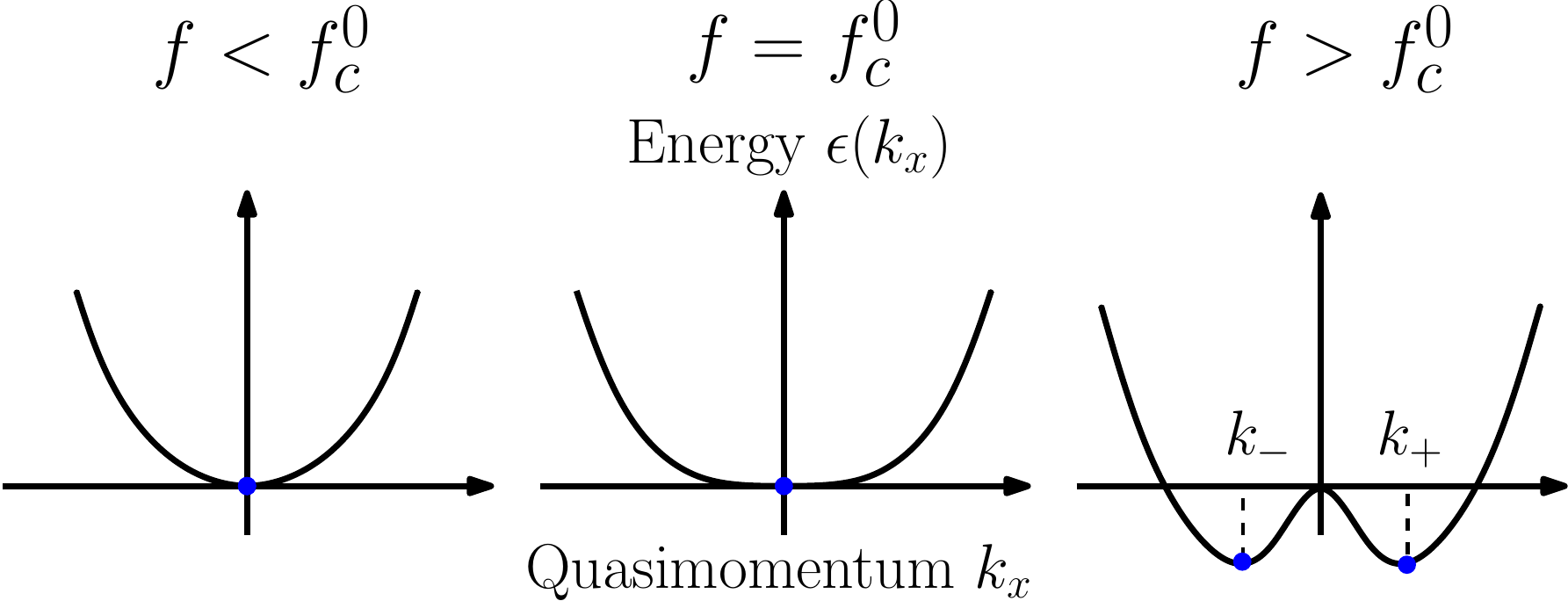}
\caption{Schematic of single-particle dispersion $\epsilon(k_x)$, changing from single minimum to double minima. An Ising type $Z_2$ quantum phase transition can be driven by shaking a lattice above a critical amplitude $f^0_\text{c}$.\label{Ising} }
\end{figure}

These systems share the following common feature. Let us consider a single particle energy-momentum dispersion along one spatial direction, say, $\epsilon(k_x)$ along $\hat{x}$. As schematically illustrated in Fig. \ref{Ising}, initially, $\epsilon(k_x)$ is a quadratic function around its unique minimum at $k_x=0$. In this regime, bosons condense into $k_x=0$ state and form a normal superfluid. As one changes a tunable parameter, at a critical point, $\epsilon(k_x)$ becomes a quartic function around $k_x=0$ and
across this point, $\epsilon(k_x)$ will display two degenerate minima at $k_\pm$. In this regime, without loss of generality, one should assume the condensate wave function as a superposition of $\varphi(k_{+})$ and $\varphi(k_{-})$, and it is up to the interaction between bosons to determine the superposition coefficients. There exist a class of systems where with weak interaction the condensate wave function will favor either purely $\varphi(k_+)$ or purely $\varphi(k_{-})$, and therefore the superfluid will break the $Z_2$ symmetry.

To bring out the novel physics of this quantum phase transition, in this letter we show that interactions can strongly modify the above picture. Our methods include both mean-field theory and a low-energy effective theory approach. This effective theory also allows us to treat both $U(1)$ and $Z_2$ symmetry breaking in a uniform framework and beyond mean-field level by perturbative renormalization group method. With these two methods, we have reached the following two results:

1) The location of the normal superfluid (SF) to superfluid that breaks an additional $Z_2$ symmetry ($Z_2$ SF) quantum critical point has a strong dependence on the interaction between particles. Bosons can condense to momentum state with high or locally maximum kinetic energy near the quantum critical point.

2) There exists a quantum tricritical point between Mott insulator (MI), SF and $Z_2$ SF phases. Interactions between atoms dictate the universal behavior and yield new universal critical exponents.

\textit{Shaken Lattice Model.}  Here we first introduce the shaking lattice model which represents a concrete realization of the superfluid Ising transition \cite{Cheng}.
As one time-periodically modulates the relative
phase $\theta$ between two counter-propagating lasers, it will result in a time-dependent lattice
potential \cite{Chu,Arimondo}
\begin{equation}
H(t)=\frac{\hat{k}_x^{2}}{2m}+V\cos^2\left( k_{0}x+\frac{\theta \left( t\right)}{2}\right) ,
\end{equation}%
where $\theta (t)=f\cos \left( \omega t\right) $, and $f$ is the shaking amplitude, $\Delta x=f/(2k_0)$ is the maximum lattice displacement. By employing the Bessel
function expansion, the lattice potential can be expressed as
\begin{equation}
\frac{V}{2}\sum_{n=-\infty }^{\infty }i^{n}J_{n}(f)\frac{%
e^{i2k_{0}x}+(-1)^{n}e^{-i2k_{0}x}}{2}e^{in\omega t}.  \label{expanding}
\end{equation}%
The $n=0$ term gives rise to a static lattice potential $VJ_{0}\left( f\right) \cos^2(k_{0}x)$, which gives a static band structure $\varepsilon _{\lambda }\left( k_x\right) $ and the Bloch wave function $\varphi
_{\lambda ,k_x}(x)$. ($\lambda $ is the band index and $k_x$ is
the quasi-momentum.)

\begin{figure}[bp]
\includegraphics[width=3.2in]
{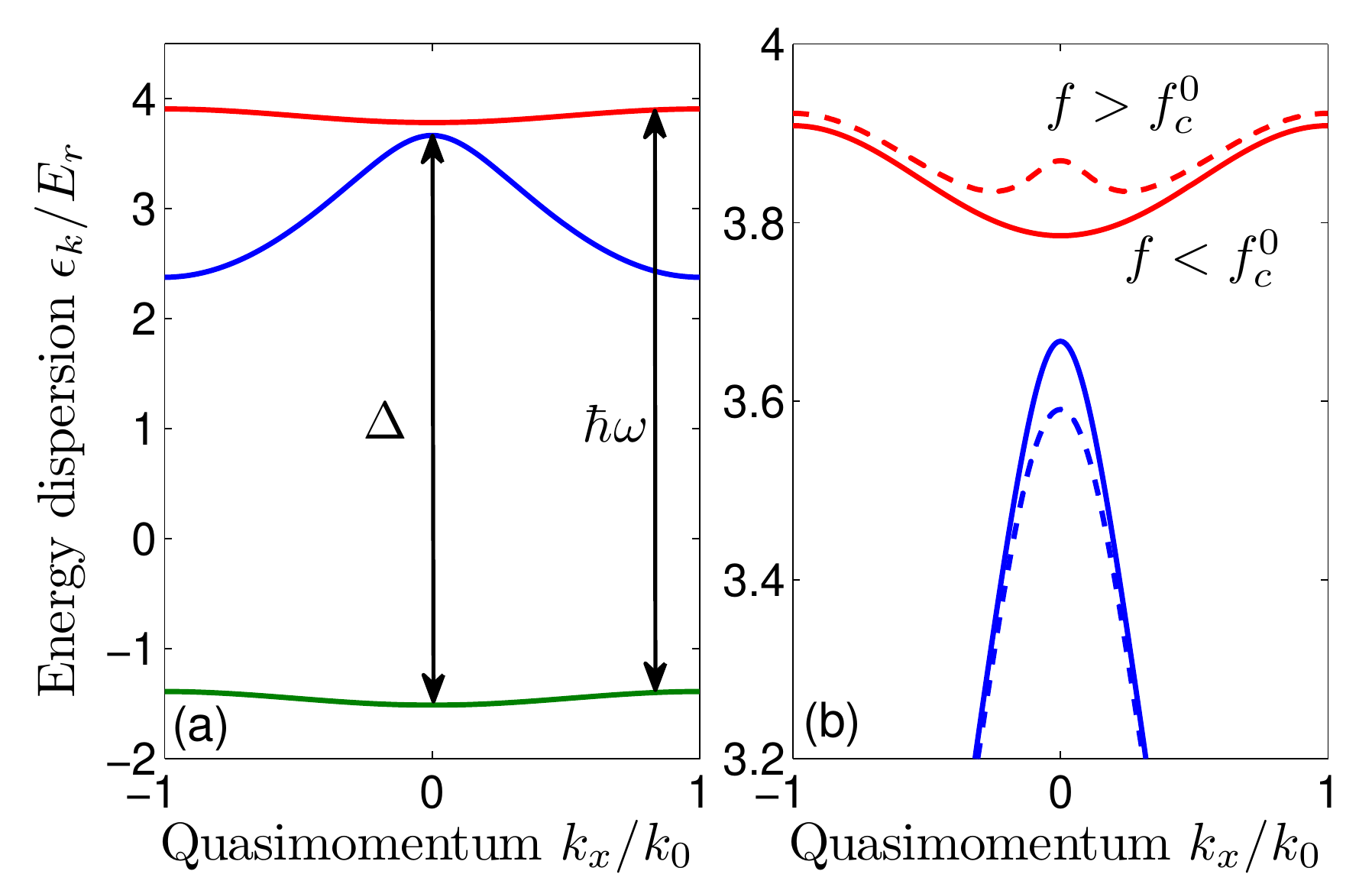}
\caption{Band structure of a shaken lattice.  (a) Band structure before shaking. The red solid line on the top is the dressed $s$-band with energy shifted by a phonon energy $\hbar\protect\omega$. (b)
Solid line is the dispersion for small shaking amplitude $f<f^0_\text{c}$ and dashed line is the dispersion for  large shaking
amplitude $f>f^0_\text{c}$. Energy is plotted in unit of lattice recoil energy $E_\text{r}=\hbar^2 k^2_0/(2m)$ }
\label{non-int}
\end{figure}

Denoting $\Delta$ as the separation between bottom of $s$-band and top of $p$-band, we consider the experimental situation $\omega \gtrsim \Delta$, as shown in Fig. \ref{non-int}(a). Here we only need to keep the most dominant $n=\pm 1$ processes in Eq. \ref{expanding} and only $s$- and $p_x$- bands, since all $|n|>1$ processes and higher bands will be generically off-resonance \cite{higher}. The time-dependent potential is now given by $V(t)=-VJ_{1}\left( f\right) \sin \left( 2k_{0}x\right) \cos \left( \omega
t\right)$,
which couples $s$- and $p_x$-bands. In the two-band bases, and upon a rotating wave approximation, it is straightforward to show that the eigen-energies are given by
\begin{align}
\epsilon _{\pm
}\left( k_x\right) =A_{k_x}/2\pm \sqrt{\Delta _{k_x}^{2}/4+\left\vert \Omega
_{k_x}\right\vert ^{2}}, \label{Ek}
\end{align}
where $\Omega _{k_x}=-VJ_{1}\left( f\right) \left\langle \varphi
_{p,k_x}\right\vert \sin \left( 2k_{0}x\right) \left\vert \varphi
_{s,k_x}\right\rangle/2$, $A_{k_x}=\varepsilon _{p}\left( k_x\right)
+\varepsilon _{s}\left( k_x\right) +\omega $, $\Delta _{k_x}=\varepsilon
_{p}\left( k_x\right) -\varepsilon _{s}\left( k_x\right) -\omega $. Two eigen wave functions are denoted by $\varphi _{+,k_x}(x)$ and $\varphi _{-,k_x}(x)$, respectively.

\begin{figure}[bp]
\includegraphics[width=3.5in]
{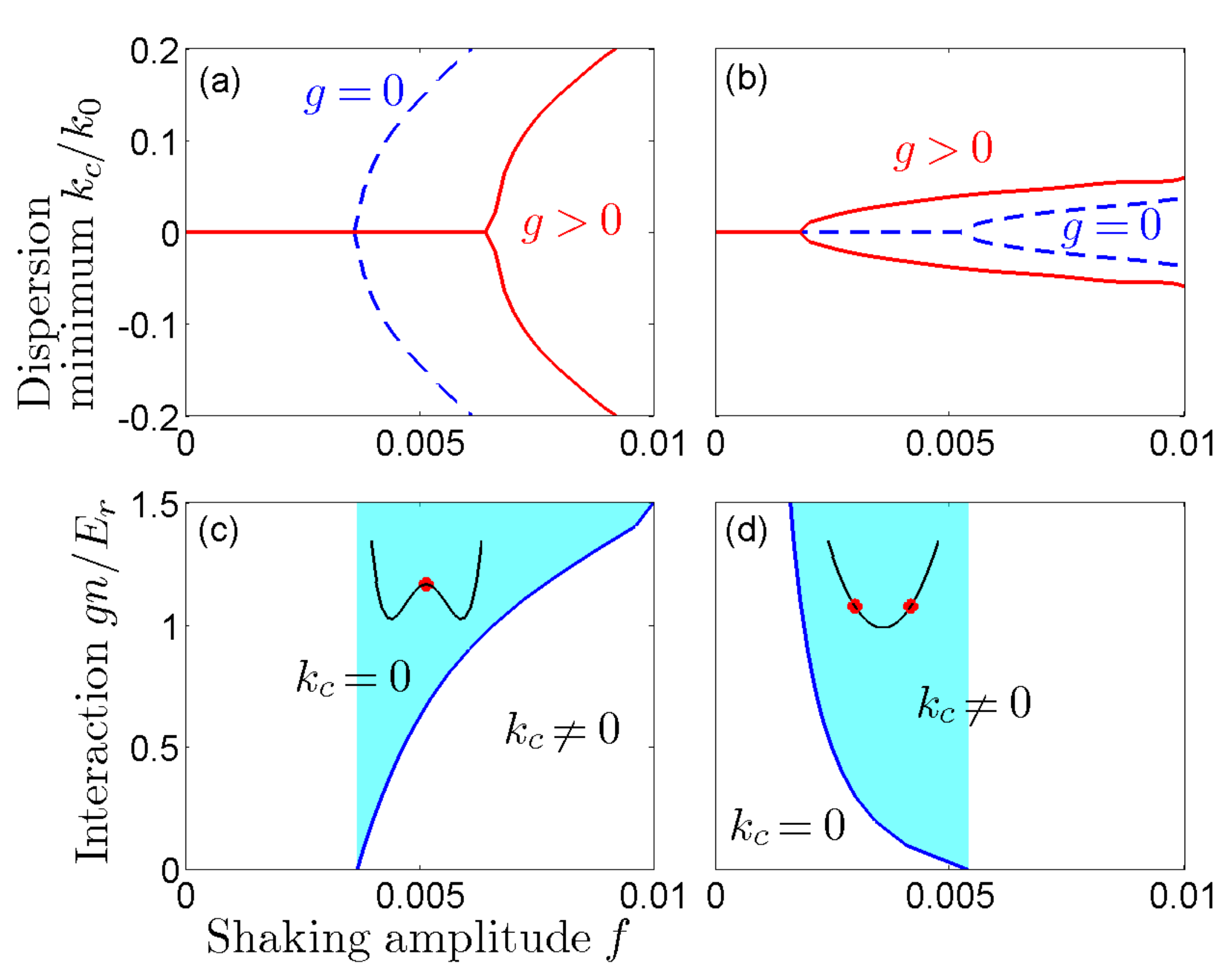}
\caption{Interaction shifts of SF-$Z_2$ SF quantum critical point. (a,c) Deep lattice with $V/E_\text{r}=16$ and $\hbar\omega/E_\text{r}=7.1$; (b,d) Shallow lattice with $V/E_\text{r}=4$ and $\hbar\omega/E_\text{r}=4.4$. (a and b) Condensate momentum $k_{\text{c}}$ as a function of $f$. Blue dashed line is for
non-interacting and red solid line is for interacting case with $gn/E_\text{r}=1$. (c and d)
Interaction($gn$)-shaking amplitude($f$) phase diagram for a fixed frequency
$\protect\omega$. Blue shaded areas show regions where atoms condense to states with finite kinetic energy.  }
\label{transition}
\end{figure}

In experiment, if one adiabatically turns on the shaking, bosons will remain in the $%
\epsilon _{+}(k_x)$ band since it is adiabatically connected to the $s$%
-band as $f\rightarrow 0$. We show in Fig. \ref{non-int}(b) that there exists a critical shaking amplitude $f^0_\text{c}$,
across which $\epsilon _{+}(k_x)$ exhibits a transition from single minimum at
zero-momentum to double minima at finite momentum $\pm k_\text{min}$. For $f>f^0_\text{c}$, without loss of generality, we can assume the condensate wave function to be a linear
superposition as $\Psi(x)=\sin\alpha \psi_{+,k_\text{min}}(x)+\cos\alpha%
\psi_{+,-k_\text{min}}(x)$. Whereas in this case the interaction energy is always minimized by choosing $\alpha$ equalling to zero or $\pi/2$ \cite{stripe}. That is to say, the condensate will break the $Z_2$ symmetry across $f^0_\text{c}$. In fact, such a transition, as well as domain wall formation in the symmetry breaking phase, has been observed in a recent experiment \cite{Cheng}.

\textit{Quantum Critical Point in an Interacting System.} Since now bosons all condense in a single momentum state, at mean-field level the interaction energy can be simplified as \cite{supp}
\begin{align}
\epsilon_\text{int}(k_x)=U^{ss}_{k_x}n^2_{s,k_x}+4U^{sp}_{k_x}n_{s,k_x}n_{p,k_x}+U^{pp}_{k_x}n^2_{p,k_x}, \label{Eint}
\end{align}
where $U_{k_x}^{\lambda ^{\prime }\lambda }=g\int dx|\varphi _{\lambda
^{\prime },k_x}|^{2}|\varphi _{\lambda ,k_x}|^{2}$, $g$ is the interaction constant, $\lambda$ and $\lambda^\prime$ denote $s$ or $p_x$. With Eq. \ref{Ek} and Eq. \ref{Eint}, the total energy of condensate is written as $\epsilon(k_x)=\epsilon_{+}(k_x)+\epsilon_{\text{int}}(k_x)$. By minimizing $\epsilon(k_x)$ with respect to $k_x$, one can determine the condensate momentum $k_\text{c}$, and further determine the critical point $f_\text{c}$ for the superfluid Ising transition when $k_\text{c}$ changes from zero to finite.

\begin{figure}[tbp]
\includegraphics[width=2.0in]
{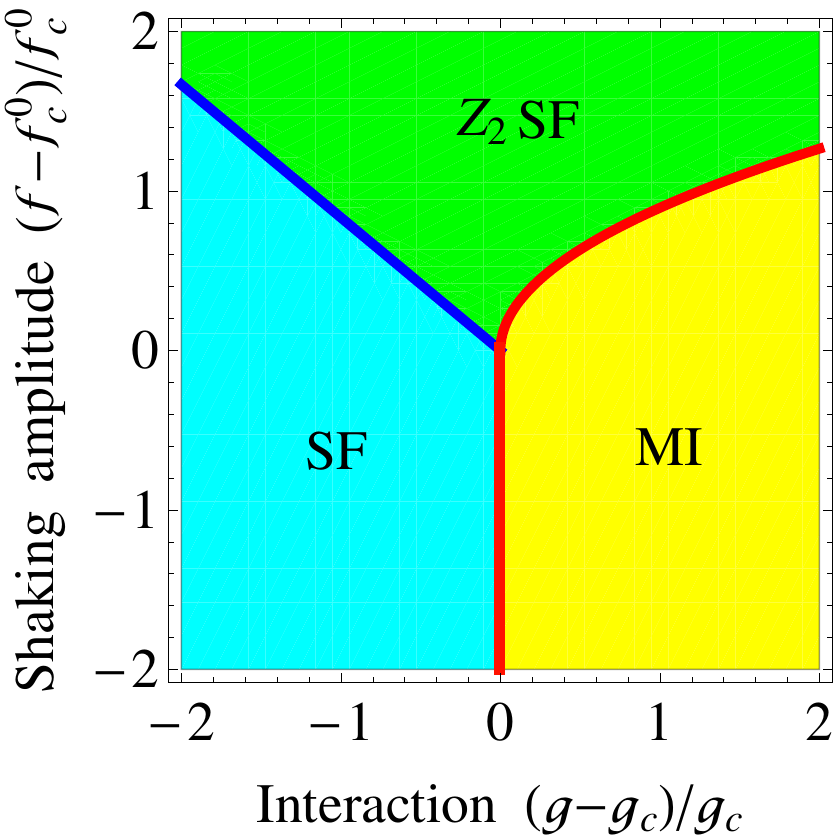}
\caption{Tricritical point of interacting bosons in a shaken lattice. Three phases are Mott insulator phase(MI), normal superfluid (SF) phase and superfluid phase that breaks $Z_2$ symmetry ($Z_2$ SF). $g_\text{c}$ defines critical interaction strength for normal superfluid to Mott insulator transition, and $f^0_\text{c}$ is the critical shaking amplitude for single particle dispersion.}
\label{phase_diagram}
\end{figure}

Our findings is shown in Fig. \ref{transition}. Because the mixing between $s$- and $p_x$-band is stronger for smaller momentum. For deep lattice in the tight binding limit (Fig. \ref{transition}(a) and (c)), the dominant contribution to $U^{\lambda^\prime\lambda}_{k_x}$ is the onsite interaction of localized Wannier states.  Since the Wannier wave function for $p_x$-band is more extended, the repulsive interaction energy has a minimum at $k_x=0$.
Therefore, in the shaded area of Fig. \ref{transition}(b), even when the kinetic energy already
exhibits double-minimum, by including interaction energy
the total energy still possesses a unique minimum at zero-momentum. In another word, in this regime, bosons are condensed into the local maximum of single particle kinetic energy, as shown in Fig. \ref{transition}(c). 

In contrast, for shallow lattice where the Bloch wave function behaves like plane waves, around $k_x\approx 0$, for $s$-band the Bloch wave function is nearly a constant,  whereas for $p_x$-band the Bloch wave function behaves like $\sin(2k_0x)$, which has stronger spatial modulation. Thus, the repulsive interaction is enhanced as $p_x$-component increases, and the interaction energy displays a local maximum at zero-momentum. Therefore, in the shaded area of Fig. \ref{transition}(d), bosons are condensed into finite momentum state which is not kinetic energy minimum state. As opposite to the tight binding limit, $f_\text{c}$ decreases as repulsive interaction increases.

This phenomenon is quite intriguing since it invalids the conventional notion that bosons always condense into its single particle energy minimum. This happens when the self-energy correction due to interactions has strong momentum dependence. So far this effect has been discussed by Li {\it et al} in spin-orbit coupling system (system (a)) \cite{Stringari}. Whereas in system (a) this shift is relatively weak because the interaction constants between different spin states are very close, in particular, for $^{87}$Rb atoms \cite{Stringari}. In the system of shaking lattice, this is due to the difference of interaction constants between different bands, such as $U^{ss}_{k_x}$, $U^{sp}_{k_x}$ and $U^{pp}_{k_x}$, and their difference is very large because of the different behaviors of Bloch wave functions. Therefore this effect is much more profound and is much easier to be observed experimentally.

\textit{Effective Theory Approach.} Next we shall go beyond the microscopic theory and present a general low-energy effective theory to describe both the superfluid and Ising transition. This effective field theory should capture two key ingredients from microscopic physics as discussed above: i) the kinetic energy expanded in term of small $k_x$ is given by $ak^2_x+bk^4_x$ where $a$ can change sign; and ii) the interaction term contains both a constant term and a term proportional to $k^2_x$. Considering a $d$-dimensional lattice with modulation along $x$-direction only, its partition function is given by
\begin{align}
&\mathcal{Z}=\int \mathcal{D}[\phi^*,\phi]\exp\{\mathcal{S}[\phi^*,\phi]\}\\
&\mathcal{S}=\int d^{d}{\bf r}d\tau\left\{K_1\phi^*\partial_\tau\phi+K_2|\partial_\tau\phi|^2+\mathcal{E}[\phi^*,\phi]\right\}
\end{align}
and (by setting $b=1$)
\begin{equation}
\mathcal{E}=|\partial^2_x\phi|^2+a|\partial_x\phi|^2+\mathcal{T}+r|\phi|^2+\alpha|\phi|^4+\beta|\phi\partial_x\phi|^2, \label{Effective}
\end{equation}
where $\phi$ is the order parameter, $\mathcal{T}=|\partial_y\phi|^2$ for $d=2$ and $\mathcal{T}=|\partial_y\phi|^2+|\partial_z\phi|^2$ for $d=3$, and the parameter $\alpha$ is positive for repulsive interactions. In the microscopic model discussed above, parameter $\beta$ can be either positive or negative. Here for simplicity, we consider the case with positive $\beta$. The parameter $a$ can be controlled by tuning the single particle dispersion, which is proportional to $f-f^0_\text{c}$ from discussion above. The parameter $r\sim g-g_\text{c}$ can be tuned by changing interaction strength $g$ that drives superfluid to Mott insulator transition.

Assuming $\phi=|\phi|e^{ik_x x}$, we can rewrite $\mathcal{E}$ as
\begin{equation}
\mathcal{E}(|\phi|,k_x)=(k^4_x +ak^2_x)|\phi|^2+r|\phi|^2+(\alpha+\beta k^2_x)|\phi|^4.
\end{equation}
The ground state is determined by $\partial \mathcal{E}/\partial k_x=0$ and $\partial \mathcal{E}/\partial |\phi|=0$, which gives rise to three different phases: $\phi=0$ as Mott  phase; $\phi\neq 0$ and $k_x=0$ as normal SF phase; and $\phi\neq 0$ and $k_x\neq 0$ as $Z_2$ SF phase. The phase diagram is given in Fig. \ref{phase_diagram} in terms of $f$ and $g$. All three phases meet at a tricritical point at $a=r=0$, around which the interaction effect is the strongest. Hereafter we concern about the critical exponent nearby the quantum tricritical point. Our discussion can be divided into two cases:

Case \textbf{A.} No particle-hole symmetry. $K_1\neq 0$ and $K_2$-term becomes irrelevant \cite{Sachdev}. In this case, it is straightforward to show that the scaling dimension of $\phi$ is $\dim[\phi]=-(2d+7)/4$ and the critical dimension is $5/2$ \cite{supp}. For a physical system with $d=2$, the scaling dimensions of $r$, $a$ and $\alpha$ are $\dim[r]=2$, $\dim[a]=1$ and $\dim[\alpha]=1/2$, respectively, and all these three terms are relevant. This is different from conventional Bose Hubbard model with quadratic dispersion, where $\dim[\alpha]=0$ and the $\alpha$-term is marginal in two-dimension. The scaling dimension of $\beta$ is $\dim[\beta]=-1/2$, and the $\beta$-term is irrelevant. That means, in this case, although the momentum-dependent interaction plays an important role at mean-field level to shift the critical value, it does not play significant role for fluctuations beyond mean-field.

\begin{figure}[tbp]
\includegraphics[width=3.3in]
{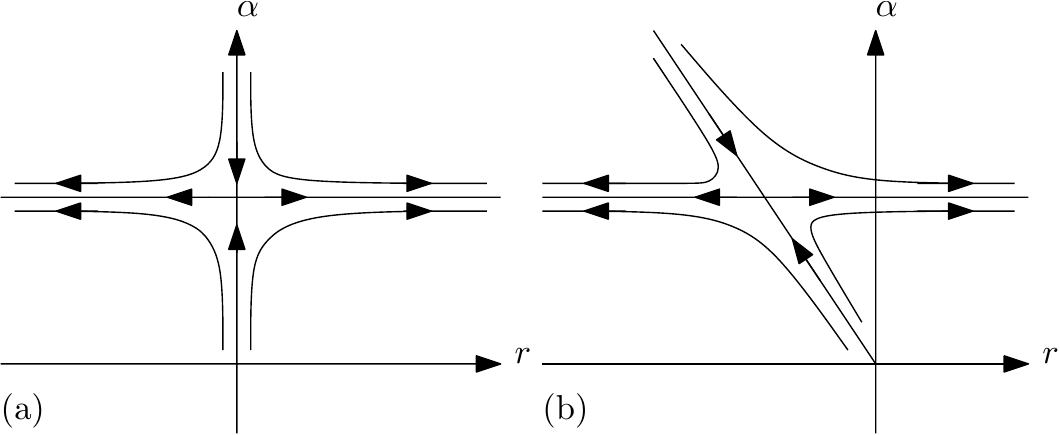}
\caption{Renormalization group flow diagram of case \textbf{A} (no particle-hole symmetry) (a) and case \textbf{B} (particle-hole symmetry) (b).}
\label{flow}
\end{figure}

In this case, the one-loop renormalizaiton group (RG) equations are derived as \cite{supp}
\begin{align}
&\frac{da}{dl}=a; \   \  \frac{dr}{dl}=2r\nonumber; \\
&\frac{d\alpha}{dl}=\epsilon\alpha-\frac{\alpha^2}{1+r}I_2(a);
\end{align}
where $\epsilon=5/2-d=1/2$, $I_2$ is a function of $a$ defined in supplementary material \cite{supp}. In addition to the Gaussion fixed point at $(a,r,\alpha)=(0,0,0)$, these RG equations give another non-Gaussion fixed point at $(a, r,\alpha)=(0,0, \epsilon/I_2(0))$. The flow diagram is shown in Fig. \ref{flow}(a). However, since in this case $r$ does not receive any correction from interaction, the critical exponent of superfluid transition still remains as its mean-field value $\nu=1/2$, as in usual Bose gas case \cite{Uzunov, Fisher}.

Case \textbf{B.} Particle-hole symmetry. $K_1=0$ \cite{Sachdev}. In this case, the scaling dimension of $\phi$ is $\dim[\phi]=-(5+2d)/4$ and the critical dimension is $7/2$. In this case, for a system in two-dimension, $\epsilon=7/2-d=3/2$. Since $\epsilon>1$ it is not accurate to treat the system by means of perturbative expansion \cite{2d}. For a system with $d=3$, $\epsilon=1/2$, and the scaling dimensions of different terms are $\dim[r]=2$, $\dim[a]=1$, $\dim[\alpha]=1/2$ and $\dim[\beta]=-1/2$, respectively. These are all identical to the case \text{A}. However, in this case, the one-loop RG equations read
\begin{align}
& \frac{da}{dl}=a; \  \  \frac{dr}{dl}=2r+\frac{2\alpha I_3(a)}{\sqrt{1+r}};     \nonumber\\
&\frac{d\alpha}{dl}=\epsilon\alpha-\frac{5I_3(a)\alpha^2}{2\sqrt{(1+r)^3}};
\end{align}
where $\epsilon=7/2-d=1/2$ and $I_3(a)$ is also defined in the supplementary material \cite{supp}. The key difference is that the $r$-term now receives correction from interaction. The new non-Gaussian fixed point is located at $(a,r,\alpha)=(0,-2\epsilon/5,2\epsilon/(5I_3(0)))$, and the flow diagram is shown in Fig. \ref{flow}(b). More importantly, the critical exponent of superfluid transition $\nu=1/(2-2\epsilon/5)=5/9$ is now different from the mean-field value \cite{supp}. This is also different from conventional bosons with $k^2$ dispersion with $K_1=0$, which belongs to the class of $O(2)$ rotor model. In this sense, it represents a new type of critical behavior.

We note that in both cases we have $\epsilon=1/2$, which is because the quartic dispersion gives rise to a fractional critical dimension. Whereas in many systems where the critical dimension is usually an integer, $\epsilon$ is at least equal to one for a physical system below critical dimension. Thus, $\epsilon$-expansion is expected to work more accurately in our cases. Previously, for conventional Bose Hubbard model the critical exponent $\nu=1/2$ has been measured with \textit{in-situ} density measurements \cite{Cheng2}. In the system of shaking lattice, one can tune the interaction to the vicinity of Mott transition, tune the chemical potential to the particle-hole symmetric point, and tune the band dispersion by shaking to the vicinity of Ising transition. Thus, case \textbf{B} can be realized and with the same \textit{in-situ} method, this new critical phenomenon predicted here can be experimentally verified.

{\it Acknowledgment}: We thank Xiaoliang Qi, Fa Wang and Hong Yao for helpful discussions. This work is supported by Tsinghua University Initiative Scientific Research Program(HZ), NSFC Grant No. 11004118 (HZ) and No. 11174176(HZ), and NKBRSFC under Grant No. 2011CB921500(HZ), NSF MRSEC (DMR-0820054)(CC), NSF Grant No. PHY- 0747907 (CC) and ARO-MURI No. 63834-PH-MUR (CC).

\begin{widetext}

\section{Supplementary Materials}

\subsection{The band structure and the interaction energy in the shaking lattice%
}

By keeping only $n=0,\pm 1$ terms, the single-particle Hamiltonian in Eq. (1)
can be separated into two parts: $H\left( t\right) =H_{0}+V\left( t\right) $,%
\begin{eqnarray}
H_{0} &=&\frac{\hat{k}_{x}^{2}}{2m}+VJ_{0}\left( f\right) \cos ^{2}\left(
k_{0}x\right) , \\
V(t) &=&-VJ_{1}\left( f\right) \sin \left( 2k_{0}x\right) \cos \left( \omega
t\right) .
\end{eqnarray}%
First part is the time-independent part, which gives a static band structure
as
\begin{equation}
H_{0}=\sum_{\lambda ,k_{x}}\varepsilon _{\lambda }\left( k_{x}\right)
\left\vert \varphi _{\lambda ,k_{x}}\right\rangle \left\langle \varphi
_{\lambda ,k_{x}}\right\vert ,  \label{H0}
\end{equation}%
where $\left\vert \varphi _{\lambda ,k_{x}}\right\rangle $ is the Bloch
function, $\lambda $ is the band index and $k_{x}$ is the quasi-momentum.
Expanding $V\left( t\right) $ in the Bloch basis, one obtains%
\begin{equation}
V\left( t\right) =-VJ_{1}\left( f\right) \cos \left( \omega t\right)
\sum_{\lambda ^{\prime }\lambda ,k_{x}^{\prime }k_{x}}\left\vert \varphi
_{\lambda ^{\prime },k_{x}^{\prime }}\right\rangle \left\langle \varphi
_{\lambda ^{\prime },k_{x}^{\prime }}\right\vert \sin \left( 2k_{0}x\right)
\left\vert \varphi _{\lambda ,k_{x}}\right\rangle \left\langle \varphi
_{\lambda ,k_{x}}\right\vert .
\end{equation}%
For the momentum transferred by $V\left( t\right) $ is $2k_{0}$, which is
just the reciprocal lattice vector, one can prove that the matrix elements
of $V\left( t\right) $ is propotianal to $\delta _{k_{x}^{\prime }k_{x}}$.
So the Hamiltonian can be simplified into:%
\begin{eqnarray}
H\left( t\right)  &=&\sum_{k_{x}}\left[ \varepsilon _{s}\left( k_{x}\right)
\left\vert \varphi _{s,k_{x}}\right\rangle \left\langle \varphi
_{s,k_{x}}\right\vert +\varepsilon _{p}\left( k_{x}\right) \left\vert
\varphi _{p,k_{x}}\right\rangle \left\langle \varphi _{p,k_{x}}\right\vert %
\right]   \notag \\
&&-VJ_{1}\left( f\right) \cos \left( \omega t\right) \sum_{k_{x}}\left\vert
\varphi _{p,k_{x}}\right\rangle \left\langle \varphi _{p,k_{x}}\right\vert
\sin \left( 2k_{0}x\right) \left\vert \varphi _{s,k_{x}}\right\rangle
\left\langle \varphi _{s,k_{x}}\right\vert +h.c.
\end{eqnarray}
One notes that Hamiltonian is diagonal in the momentum space, $H\left(
t\right) =\sum_{k_{x}}H\left( k_{x},t\right) $. And $H\left( k_{x},t\right) $
can be rewritten into a matrix form%
\begin{equation}
H\left( k_{x},t\right) =\left(
\begin{array}{cc}
\varepsilon _{p}\left( k_{x}\right)  & 2\Omega _{k_{x}}\cos \left( \omega
t\right)  \\
2\Omega _{k_{x}}^{\ast }\cos \left( \omega t\right)  & \varepsilon
_{s}\left( k_{x}\right)
\end{array}%
\right) .
\end{equation}%
where $\Omega _{k_{x}}=-\frac{1}{2}VJ_{1}\left( f\right) \left\langle
\varphi _{p,k}\right\vert \sin \left( 2k_{0}x\right) \left\vert \varphi
_{s,k}\right\rangle $. Making a unitary transformation,
\begin{equation}
U\left( t\right) =\left(
\begin{array}{cc}
1 & 0 \\
0 & e^{i\omega t}%
\end{array}%
\right) ,
\end{equation}%
one obtains the Hamiltonian in the rotational frame,%
\begin{equation}
H^{\prime }\left( k_{x},t\right) =\left(
\begin{array}{cc}
\varepsilon _{p}\left( k_{x}\right)  & \Omega _{k_{x}}\left( 1+e^{2i\omega
t}\right)  \\
\Omega _{k_{x}}^{\ast }\left( 1+e^{-2i\omega t}\right)  & \varepsilon
_{s}\left( k_{x}\right) +\omega
\end{array}%
\right) .  \notag
\end{equation}%
Omitting the high frequency oscillation terms (this is so-called rotational
wave approximation), one obtains the time-independent Hamiltonian,%
\begin{equation*}
H^{\prime }\left( k_{x}\right) =\left(
\begin{array}{cc}
\varepsilon _{p}\left( k\right)  & \Omega _{k} \\
\Omega _{k}^{\ast } & \varepsilon _{s}\left( k\right) +\omega
\end{array}%
\right) .
\end{equation*}%
Diagoanlizing $H^{\prime }\left( k_{x}\right) $, one obtains the energy
bands as
\begin{equation}
\epsilon _{\pm }\left( k_{x}\right) =A_{k_{x}}/2\pm \sqrt{\Delta
_{k_{x}}^{2}/4+\left\vert \Omega _{k_{x}}\right\vert ^{2}},
\end{equation}%
where $A_{k_{x}}=\varepsilon _{p}\left( k_{x}\right) +\varepsilon _{s}\left(
k_{x}\right) +\omega $, $\Delta _{k_{x}}=\varepsilon _{p}\left( k_{x}\right)
-\varepsilon _{s}\left( k_{x}\right) -\omega $. The correspoding two eigen
wave functions are%
\begin{eqnarray}
\varphi _{+,k_{x}}\left( x\right)  &=&c_{+,s}\left( k_{x}\right) \varphi
_{s,k_{x}}\left( x\right) +c_{+,p}\left( k_{x}\right) \varphi
_{p,k_{x}}\left( x\right) ,  \label{wf_up} \\
\varphi _{-,k_{x}}\left( x\right)  &=&c_{-,s}\left( k_{x}\right) \varphi
_{s,k_{x}}\left( x\right) +c_{-,p}\left( k_{x}\right) \varphi
_{p,k_{x}}\left( x\right) ,
\end{eqnarray}%
where $c_{\pm ,\lambda }\left( k_{x}\right) $ is the combination coefficient
obtained from the Diagoanlization of $H^{\prime }\left( k_{x}\right) $.
Transforming the wave funtions back to the laboratorial frame, we obtain
\begin{eqnarray}
\varphi _{+,k_{x}}\left( x,t\right)  &=&e^{-i\omega t}c_{+,s}\left(
k_{x}\right) \varphi _{s,k_{x}}\left( x\right) +c_{+,p}\left( k_{x}\right)
\varphi _{p,k_{x}}\left( x\right) , \\
\varphi _{-,k_{x}}\left( x,t\right)  &=&e^{-i\omega t}c_{-,s}\left(
k_{x}\right) \varphi _{s,k_{x}}\left( x\right) +c_{-,p}\left( k_{x}\right)
\varphi _{p,k_{x}}\left( x\right) .
\end{eqnarray}%
Considering the Bose condensation in the upper band, the time-average
interaction energy is a funtion of condensate quasi-momentum,%
\begin{eqnarray*}
\epsilon _{\mathrm{int}}\left( k_{x}\right)  &=&\frac{1}{T}%
g\int_{0}^{T}dt\int dx\left\vert \varphi _{+,k_{x}}\left( x,t\right)
\right\vert ^{4} \\
&=&U_{k_{x}}^{ss}\left\vert c_{+,s}\left( k_{x}\right) \right\vert
^{4}+4U_{k_{x}}^{sp}\left\vert c_{+,s}\left( k_{x}\right) \right\vert
^{2}\left\vert c_{+,s}\left( k_{x}\right) \right\vert
^{2}+U_{k_{x}}^{pp}\left\vert c_{+,p}\left( k_{x}\right) \right\vert ^{4}
\end{eqnarray*}%
where $U_{k_{x}}^{\lambda ^{\prime }\lambda }=g\int dx\left\vert \varphi
_{\lambda ^{\prime },k_{x}}\left( x\right) \right\vert ^{2}\left\vert
\varphi _{\lambda ,k_{x}}\left( x\right) \right\vert ^{2}$, $g$ is the
interaction constant, $\lambda ^{\prime }$ and $\lambda $ denote $s$ or $p$.
$|c_{+,\text{s}}\left( k_{x}\right)|^2$ and $|c_{+,\text{p}}\left( k_{x}\right)|^2$ are denoted by $n_{\text{s},k_x}$ and $n_{\text{p},k_x}$ in Eq. 5 of the main text , respectively.

\subsection{The mean-field study of the low energy effective theory}

A low energy effective theory that describes both superfluid and Ising
transitions can be constructed as Eq. (6)-(8) in the main text. In the
momentum space the energy density can be written as
\begin{eqnarray}
\mathcal{E}=(k_c^4+ak_c^2)|\phi|^2+r|\phi|^2+(\alpha+\beta k_c^2)|\phi|^4,
\end{eqnarray}
where we ignore the $k_y$ and $k_z$ dependent terms since the energy minima
would be always located at $k_y=k_z=0$. The ground state is determined by
minimizing the energy density as the following:
\begin{eqnarray}
&& \frac{\partial\mathcal{E}}{\partial k_c}=0\Rightarrow[(2k_c^2+a)+\beta|%
\phi|^2]k_c|\phi|^2=0,\cr&& \frac{\partial \mathcal{E}}{\partial |\phi|}%
=0\Rightarrow [(k_c^4+ak_c^2+r)+2(\alpha+\beta k_c^2)|\phi|^2]|\phi|=0.
\end{eqnarray}
We study the energy minimum in four regions of the parameter space, where $%
\alpha$ and $\beta$ are always positive.

\begin{enumerate}
\item In the region $a>0$ and $r>0$ neither of equations $%
(2k_c^2+a)+\beta|\phi|^2=0$ and $(k_c^4+ak_c^2+r)+2(\alpha+\beta
k_c^2)|\phi|^2=0$ has solutions. The energy minimum is at $|\phi|=0$.

\item In the region $a>0$ and $r<0$ equation $(2k_c^2+a)+\beta|\phi|^2=0$
doesn't have any solution. Then we take $k_c=0$ and solve equation $%
(k_c^4+ak_c^2+r)+2(\alpha+\beta k_c^2)|\phi|^2=0$. The energy minimum is
located at $k_c=0, ~|\phi|=-\frac{r}{2\alpha}$.

\item In the region $a<0$ and $r>0$ both equations $(2k_c^2+a)+\beta|%
\phi|^2=0$ and $(k_c^4+ak_c^2+r)+2(\alpha+\beta k_c^2)|\phi|^2=0$ can have
solutions, which are
\begin{eqnarray}
&&|\phi|^2=-\frac{2k_c^2+a}{\beta},\cr && |\phi|^2=-\frac{k_c^4+ak_c^2+r}{%
2(\alpha+\beta k_c^2)}.
\end{eqnarray}
Then $k_c^2$ can be solved from the equation
\begin{eqnarray}
-\frac{2k_c^2+a}{\beta}=-\frac{k_c^4+ak_c^2+r}{2(\alpha+\beta k_c^2)}.
\end{eqnarray}
The above equation has two roots as
\begin{eqnarray}
k_c^2=\frac{-(4r\alpha+a\beta)\pm\sqrt{(4r\alpha+a\beta)^2-12\beta(2a\alpha-%
\beta r)}}{6\beta}.
\end{eqnarray}
Here we ignore the negative root since $k_c^2>0$.

To guarantee that the solution of Eq. (5) is valid two restriction
conditions should be satisfied as the following:

\begin{enumerate}
\item We have
\begin{eqnarray}
-\frac{2k_c^2+a}{\beta}>0
\end{eqnarray}
since $|\phi|^2>0$ in the Eq. (3).

\item We have
\begin{eqnarray}
-(4r\alpha+a\beta)+\sqrt{(4r\alpha+a\beta)^2-12\beta(2a\alpha-\beta r)}>0
\end{eqnarray}
since $k_c^2>0$ in the Eq. (5).
\end{enumerate}

In the region $a<0$ and $r>0$ it's straight forward to check that the
condition (b) is always satisfied, while the condition (a) generate an upper
bound. We can obtain this boundary by plugging the root of $k_c^2=\frac{%
-(4r\alpha+a\beta)+\sqrt{(4r\alpha+a\beta)^2-12\beta(2a\alpha-\beta r)}}{%
6\beta}$ into the condition (a). Then the upper bound is
\begin{eqnarray}
r<\frac{a^2}{4}.
\end{eqnarray}
Below the boundary of Eq. (8) we have the energy minimum at
\begin{eqnarray}
&&k_c^2=\frac{-(4r\alpha+a\beta)+\sqrt{(4r\alpha+a\beta)^2-12\beta(2a\alpha-%
\beta r)}}{6\beta}, \cr && |\phi|^2=\frac{-4r\alpha+2a\beta+\sqrt{%
(4r\alpha+a\beta)^2-12\beta(2a\alpha-\beta r)}}{3\beta^2}.
\end{eqnarray}

\item In the region $a<0$ and $r<0$ the energy minimum can be also located
as Eq. (9). Here the restriction condition (b) generates a lower bound as
\begin{eqnarray}
r>\frac{2\alpha}{\beta} a.
\end{eqnarray}
\end{enumerate}

\subsection{Renormalization group analysis}

\subsubsection{The system without particle-hole symmetry}

In the system without particle-hole symmetry we have $K_1\neq0$ in Eq. (7)
of the main text. The term of $K_2$ becomes irrelevant and can be ignored.
Then the partition function in $d$-dimensions is cast as
\begin{eqnarray}
\mathcal{Z}=\int D[\phi^\ast,\phi]e^{-S[\phi^\ast,\phi]},
\end{eqnarray}
where
\begin{eqnarray}
S[\phi^\ast,\phi]=&&\int d^d \vec x d\tau \Big\{\phi^\ast(\vec
x,\tau)\partial_\tau\phi(\vec x,\tau) +|\partial_x^2\phi(\vec
x,\tau)|^2+a|\partial_x\phi(\vec x,\tau)|^2+\mathcal{T }\cr &&+r|\phi(\vec
x,\tau)|^2+\alpha|\phi(\vec x,\tau)|^4+\beta|\phi(\vec
x,\tau)\partial_x\phi(\vec x,\tau)|^2\Big\},
\end{eqnarray}
where $\mathcal{T}=|\partial_y\phi|^2$ for $d=2$ and $\mathcal{T}%
=|\partial_y\phi|^2+|\partial_z\phi|^2$ for $d=3$.

We Fourier transform $\phi(\vec x,\tau)$ as
\begin{eqnarray}
\phi(\vec x,\tau)=\int^\infty_{-\infty}\frac{d\omega}{2\pi}\int\frac{d^dk}{%
(2\pi)^2}\phi(\omega,\vec k)e^{i(\vec k\cdot\vec x-\omega\tau)}.
\end{eqnarray}
Then the action can be written in the momentum space as
\begin{eqnarray}
S=&&\int^\infty_{-\infty}\frac{d\omega}{2\pi}\int\frac{d^dk}{(2\pi)^2}
\phi^\ast(\omega,\vec k)(-i\omega+k_x^4+ak_x^2+\mathcal{T}%
_k+r)\phi(\omega,\vec k)\cr&&+\int_{\omega k}^\Lambda\Big\{(\alpha+\beta
k_{3x}k_{1x})\phi_i^\ast(\omega_4,\vec k_4)\phi_i^\ast(\omega_3,\vec
k_3)\phi_i(\omega_2,\vec k_2)\phi_i(\omega_1,\vec k_1)\Big\},
\end{eqnarray}
where $\mathcal{T}_k=k_y^2$ for $d=2$ and $\mathcal{T}=k_y^2+k_z^2$ for $d=3$%
. Here we used a short-handed notation $\int_{\omega
k}^\Lambda=\prod^4_{i=1}\int^\infty_{-\infty} \frac{d\omega_i}{2\pi}%
\int^\Lambda_0\frac{d^d k_i}{(2\pi)^2}(2\pi)^2\delta(\vec k_4+\vec k_3-\vec
k_2-\vec k_1)\cdot(2\pi)\delta(\omega_4+\omega_3-\omega_2-\omega_1)$.

Following the Wilson's approach the renormalization group transformation
involves three steps: (i) integrating out all momenta between $\Lambda/s$
and $\Lambda$, for tree level analysis just discarding the part of the
action in this momentum-shell; (ii) rescaling frequencies and the momenta as
$(\omega, k_x, k_y)\rightarrow (s^{[\omega]}\omega, s^{[k_x]}k_x, sk_y)$ so
that the cutoff in k is once again at $\pm\Lambda$; and finally (iii)
rescaling fields $\phi \rightarrow s^{[\phi]} \phi$ to keep the free-field
action $S_0$ invariant.

After we integrate out a thin momentum shell of high energy mode the limit
of $k_y$ changes from $[0, \Lambda]$ to $[0, \Lambda/s]$ and the limit of $%
k_x$ changes from $[0, \Lambda]$ to $[0, \Lambda/\sqrt{s}]$ , where $%
s\gtrapprox1$. In order to compare the action with the original one we need
to rescale the coordinate as
\begin{equation}
k_x^{\prime }=\sqrt s k_x, ~~~k_y^{\prime }=sk_y.
\end{equation}
Hence, the cutoff in $k_x$ and $k_y$ are back again at $\Lambda$. Here we
give a definition to the scaling dimension. If a quantity scales as
\begin{equation}
A^{\prime [A]}A,
\end{equation}
we call $[A]$ the scaling dimension of $A$. In this manner the scaling
dimensions of momentum $k_x$ and $k_y$ are
\begin{equation}
[k_x]=\frac{1}{2}~~~\mbox{and}~~~ [k_y]=1.
\end{equation}
A straight forward scaling analysis shows that the scaling dimensions of the
parameters are
\begin{eqnarray}
&&[\omega]=2,~~~[a]=1, ~~~[r]=2,\cr &&[\phi]=-\frac{7+2d}{4}, ~~~[\alpha]=%
\frac{5}{2}-d,~~~ [\beta]=\frac{3}{2}-d.
\end{eqnarray}
We see that upper critical dimension is $5/2$.

For a $d=2$ system the scaling dimension of $\beta$ is $-\frac{1}{2}$, which
is irrelevant. Therefore, we ignore the $\beta$ term in the following
calculations. The one-loop correction to the parameter $a$ is fully
generated by the $\beta$ term. Since the $\beta$ term is ignored we don't
have the one-loop correction to the parameter $a$. Then the flow equation of
$a$ just includes the tree-level scaling as
\begin{eqnarray}
\frac{da}{d\ell}=a.
\end{eqnarray}

The one-loop corrections to $r$ and $\alpha$ are presented in Fig. 1.
\begin{figure}[h]
\begin{center}
\includegraphics[width=7cm]{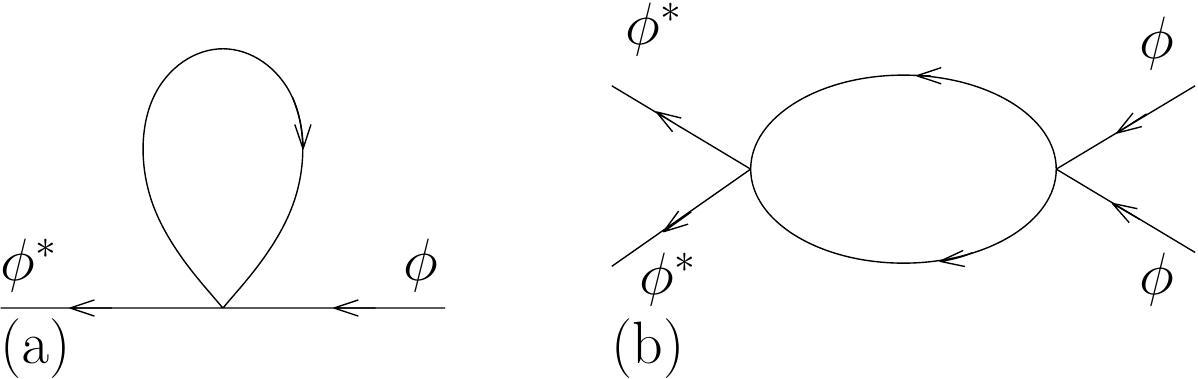}
\end{center}
\caption{The one-loop Feynman graphs contributing to the renormalization of
(a) the parameter $r$, (b) the parameter $\protect\alpha$ in systems without
particle-hole symmetry.}
\label{fig:oneloop1}
\end{figure}

By integrating out the momentum shell we get the one-loop correction to $r$
as
\begin{eqnarray}
4\alpha \int^\infty_{-\infty}\frac{d\omega}{2\pi}\int_{\mathrm{shell}}\frac{%
d^2k}{(2\pi)^2}\frac{1}{-i\omega+k_x^4+ak_x^2+k^2_y+r}.
\end{eqnarray}
The integration over the Matsubara frequency $\omega$ can be calculated by
performing a contour integration.
\begin{eqnarray}
&&\int^\infty_{-\infty}\frac{d\omega}{2\pi}\frac{1}{-i%
\omega+k_x^4+ak_x^2+k^2_y+r}\cr =&&\int_{C}\frac{dz}{2\pi i}\frac{e^{z0^+}}{%
-z+k_x^4+ak_x^2+k^2_y+r}\cr =&&-\theta(-(k_x^4+ak_x^2+k^2_y+r)),
\end{eqnarray}
where $z=i\omega$ and contour C is over the left plane. We start our RG flow
from the Gaussian fixed point, where $a=0, r=0$. Then the $\theta$ function $%
\theta(-(k_x^4+k^2_y))$ vanish since $k_x^4+k^2_y>0$. The one-loop
correction to the parameter $r$ is zero. Then the flow equation of $r$ just
includes a tree-level scaling term as
\begin{eqnarray}
\frac{dr}{d\ell}=2r.
\end{eqnarray}

The one-loop correction to the parameter $\alpha$ is
\begin{eqnarray}
-2\alpha^2\int^\infty_{-\infty}\frac{d\omega}{2\pi}\int_{\mathrm{shell}}%
\frac{d^2k}{(2\pi)^2}\frac{1}{-i\omega+k_x^4+ak_x^2+k^2_y+r}\cdot\frac{1}{%
i\omega+k_x^4+ak_x^2+k^2_y+r}.
\end{eqnarray}
The integration over the Matsubara frequency $\omega$ can be done by
performing a contour integration.
\begin{eqnarray}
&&\int^\infty_{-\infty}\frac{d\omega}{2\pi}\frac{1}{-i%
\omega+k_x^4+ak_x^2+k^2_y+r}\cdot\frac{1}{i\omega+k_x^4+ak_x^2+k^2_y+r} \cr%
=&&\frac{1}{2(k_x^4+ak_x^2+k^2_y+r)}.
\end{eqnarray}
The integration over the momentum is as the following:
\begin{eqnarray}
&& \int_{\mathrm{shell}}\frac{dk_xdk_y}{(2\pi)^2}\frac{1}{%
2(k_x^4+ak_x^2+k^2_y+r)}\cr=&&\int_{\mathrm{shell}}\frac{kdkd\theta}{(2\pi)^2%
}\frac{1}{2(k_x^4+ak_x^2+k^2_y+r)}.
\end{eqnarray}
$k_x$ scales as $k_x^{\prime }=\sqrt s k_x=e^{\frac{1}{2}\ell}k_x$, then $%
dk_x=\frac{1}{2}k_xd\ell=\frac{1}{2}k\cos\theta d\ell$. In the same manner
we have $dk_y=k\sin\theta d\ell$. Then $dk=\sqrt{(dk_x)^2+(dk_y)^2}=k\sqrt{%
\frac{1}{4}\cos^2\theta+\sin^2\theta}d\ell$. The cutoff is set as $%
k_x^4+ak_x^2+k^2_y=\Lambda^2$. $k^2$ can be solved as $k^2=\frac{%
-(\sin^2\theta+a\cos^2\theta)+\sqrt{(\sin^2\theta+a\cos^2\theta)^2+4%
\Lambda^2\cos^4\theta}}{2\cos^4\theta}$. In the following calculations we
will conveniently set $\Lambda=1$ and henceforth measure all lengths in
units of $\Lambda^{-1}$. The integration becomes
\begin{eqnarray}
&&\int_{\mathrm{shell}}\frac{kdkd\theta}{(2\pi)^2}\frac{1}{%
2(k_x^4+ak_x^2+k^2_y+r)}\cr=&&\frac{d\ell}{2(1+r)}\int^{2\pi}_0 \frac{d\theta%
}{(2\pi)^2} \frac{-(\sin^2\theta+a\cos^2\theta)+\sqrt{(\sin^2\theta+a\cos^2%
\theta)^2+4\cos^4\theta}}{2\cos^4\theta}\sqrt{\frac{1}{4}\cos^2\theta+\sin^2%
\theta}\cr=&&\frac{d\ell}{2(1+r)} \cdot I_2(a),
\end{eqnarray}
where the function $I_2(a)$ is defined as%
\begin{equation}
I_2(a)=\int^{2\pi}_0 \frac{d\theta}{(2\pi)^2} \frac{-(\sin^2\theta+a\cos^2%
\theta)+\sqrt{(\sin^2\theta+a\cos^2\theta)^2+4\cos^4\theta}}{2\cos^4\theta}%
\sqrt{\frac{1}{4}\cos^2\theta+\sin^2\theta}.
\end{equation}
After the third step of rescaling in the renormalization group
transformation the flow equation of $\alpha$ is calculated as
\begin{eqnarray}
\frac{d\alpha}{d\ell}=\epsilon\alpha-\frac{\alpha^2}{1+r}\cdot I_2(a),
\end{eqnarray}
where $\epsilon=\frac{5}{2}-d$.

Thus, we have all the flow equations as the following:
\begin{eqnarray}
&&\frac{da}{d\ell}=a,\cr&&\frac{dr}{d\ell}=2r,\cr&&\frac{d\alpha}{d\ell}%
=\epsilon\alpha-\frac{\alpha^2}{1+r}\cdot I_2(a).
\end{eqnarray}
There are two fixed points. One is the Gaussian fixed point $(a, r,
\alpha)=(0,0,0)$, the other one the Gaussian-like fixed point$%
(a^\ast,r^\ast, \alpha^\ast)=(0,0,\frac{\epsilon}{I_2(0)})$. Now we study
the structure of the the flows near the new fixed point. Defining $%
a=a^\ast+\delta a$, $r=r^\ast+\delta r$ and $\alpha=\alpha^\ast+\delta
\alpha $ yields the linearized flow equations
\begin{eqnarray}
&&\frac{d\delta a}{d\ell}=\delta a,\cr&&\frac{d\delta r}{d\ell}=2\delta r,\cr%
&&\frac{d\delta \alpha}{d\ell}=-\epsilon\delta \alpha.
\end{eqnarray}
Then the scaling exponent of $r$ is $y_r=2$. If we use $\delta=|r-r_c|$ to
define the distance to the critical point, the correlation length should
scales as $\xi\sim \delta^{-\nu}$. The scaling analysis shows that $%
\nu=1/y_r=1/2$.

\subsubsection{The system with particle-hole symmetry}

In the system with particle-hole symmetry the $K_1$ term vanishes in the Eq.
(7) of the main text. Then the partition function in $d$-dimensions is
written as
\begin{eqnarray}
\mathcal{Z}=\int D[\phi^\ast,\phi]e^{-S[\phi^\ast,\phi]},
\end{eqnarray}
where
\begin{eqnarray}
S[\phi^\ast,\phi]=&&\int d^d \vec x d\tau \Big\{|\partial_\tau\phi(\vec
x,\tau)|^2 +|\partial_x^2\phi(\vec x,\tau)|^2+a|\partial_x\phi(\vec
x,\tau)|^2+\mathcal{T }\cr &&+r|\phi(\vec x,\tau)|^2+\alpha|\phi(\vec
x,\tau)|^4+\beta|\phi(\vec x,\tau)\partial_x\phi(\vec x,\tau)|^2\Big\}.
\end{eqnarray}

A straight forward scaling analysis shows that the scaling dimensions of the
parameters are
\begin{eqnarray}
&&[k_x]=\frac{1}{2}, ~~~[k_y]=1,\cr && [\omega]=1,~~~[a]=1, ~~~[r]=2,\cr %
&&[\phi]=-\frac{5+2d}{4}, ~~~[\alpha]=\frac{7}{2}-d,~~~ [\beta]=\frac{5}{2}%
-d.
\end{eqnarray}
The upper critical dimension is $\frac{7}{2}$. In two dimensions $[\alpha]=%
\frac{3}{2}$ and $[\beta]=\frac{1}{2}$. Both of them are relevant. In three
dimensions $[\alpha]=\frac{1}{2}$ is relevant and $[\beta]=-\frac{1}{2}$ is
irrelevant.

Here we consider the $d=3$ system. In this case the $\beta$ term is
irrelevant, which will be ignored in our consideration. Then parameter $a$
received no corrections at one-loop level. The flow equation of $a$ is
\begin{equation}
\frac{da}{d\ell}=a.
\end{equation}

The Feynman graphs contributing to parameters $r$ and $\alpha$ are shown in
Fig. 2.
\begin{figure}[h]
\begin{center}
\includegraphics[width=9cm]{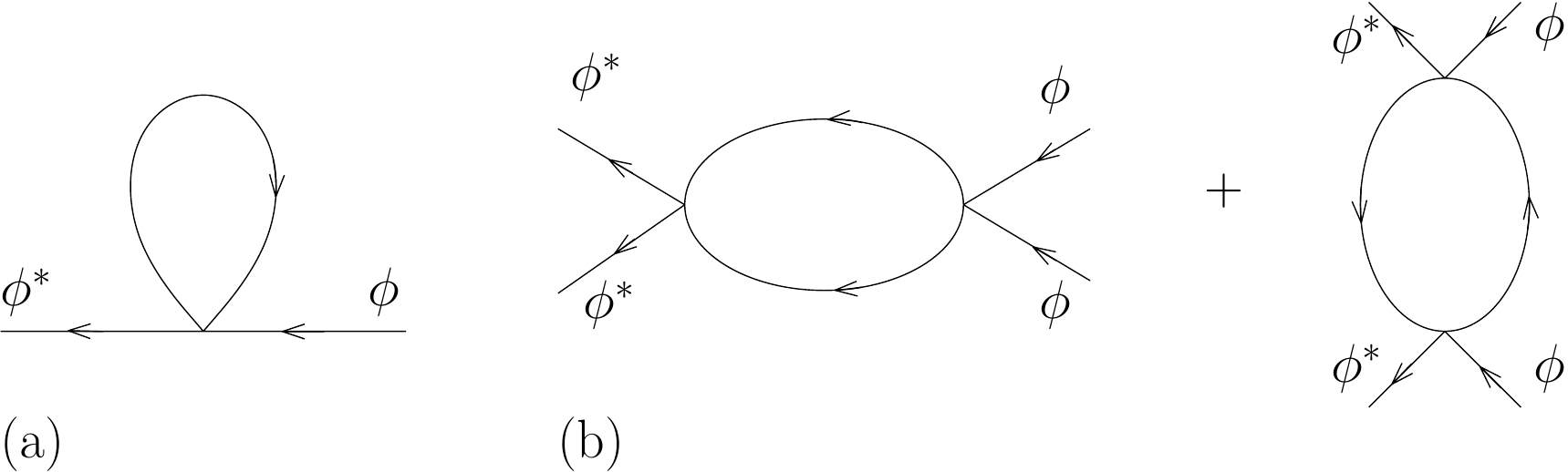}
\end{center}
\caption{The one-loop Feynman graphs contributing to the renormalization of
(a) the parameter $r$, (b) the parameter $\protect\alpha$ in systems with
particle-hole symmetry.}
\label{fig:oneloop1}
\end{figure}

The one-loop correction to $r$ is given as
\begin{eqnarray}
4\alpha \int^\infty_{-\infty}\frac{d\omega}{2\pi}\int_{\mathrm{shell}}\frac{%
d^3k}{(2\pi)^3}\frac{1}{\omega^2+k_x^4+ak_x^2+k^2_y+k_z^2+r}.
\end{eqnarray}
The integration over the Matsubara frequency $\omega$ can be calculated by
performing a contour integration.
\begin{eqnarray}
&&\int^\infty_{-\infty}\frac{d\omega}{2\pi}\frac{1}{%
\omega^2+k_x^4+ak_x^2+k^2_y+k_z^2+r}\cr =&&\int_{C}\frac{dz}{2\pi i}\frac{1}{%
\Big(-z+\sqrt{k_x^4+ak_x^2+k^2_y+k_z^2+r}\Big)\Big(z+\sqrt{%
k_x^4+ak_x^2+k^2_y+k_z^2+r}\Big)}\cr =&&\frac{1}{2\sqrt{%
k_x^4+ak_x^2+k^2_y+k_z^2+r}},
\end{eqnarray}
where $z=i\omega$ and contour C is over the left plane. Analogous to the
procedure in Eq. (25)-(28) the momentum shell integration can be performed
as
\begin{eqnarray}
&& \int_{\mathrm{shell}}\frac{dk_xdk_ydk_z}{(2\pi)^3}\frac{1}{2\sqrt{%
k_x^4+ak_x^2+k^2_y+k_z^2+r}}\cr=&&\frac{d\ell}{2\sqrt{1+r}}\cdot I_3(a),
\end{eqnarray}
where the function $I_3$ is defined as
\begin{eqnarray}
I_3(a)=&& \int^{2\pi}_0 \frac{d\theta}{(2\pi)^2} \Bigg(\frac{%
-(\sin^2\theta+a\cos^2\theta)+\sqrt{(\sin^2\theta+a\cos^2\theta)^2
+4\cos^4\theta}}{2\cos^4\theta}\Bigg)^\frac{3}{2}\cdot\sin\theta \sqrt{\frac{%
1}{4}\cos^2\theta+\sin^2\theta}.
\end{eqnarray}
Then we have the flow equation of $r$ as
\begin{eqnarray}
\frac{dr}{d\ell}=2r+\frac{2\alpha}{\sqrt{1+\alpha}}\cdot I_3(a).
\end{eqnarray}

The one-loop correction to the parameter $\alpha$ is
\begin{eqnarray}
-10\alpha^2\int^\infty_{-\infty}\frac{d\omega}{2\pi}\int_{\mathrm{shell}}%
\frac{d^3k}{(2\pi)^3}\frac{1}{(\omega^2+k_x^4+ak_x^2+k^2_y+k_z^2+r)^2}.
\end{eqnarray}
The integration over the Matsubara frequency $\omega$ can be done by
performing a contour integration.
\begin{eqnarray}
&&\int^\infty_{-\infty}\frac{d\omega}{2\pi}\frac{1}{%
(\omega^2+k_x^4+ak_x^2+k^2_y+k_z^2+r)^2} \cr=&&\int_{C}\frac{dz}{2\pi i}%
\frac{1}{\Big(-z+\sqrt{k_x^4+ak_x^2+k^2_y+k_z^2+r}\Big)^2 \Big(z+\sqrt{%
k_x^4+ak_x^2+k^2_y+k_z^2+r}\Big)^2}\cr =&&\frac{1}{%
4(k_x^4+ak_x^2+k^2_y+k_z^2+r)^\frac{3}{2}}.
\end{eqnarray}
Then it's straight forward to obtain the flow equation of $\alpha$ as
\begin{eqnarray}
\frac{d\alpha}{d\ell}=\epsilon\alpha-\frac{5}{2}\frac{I_3(a)}{(1+r)^\frac{3}{%
2}}\alpha^2.
\end{eqnarray}

Then all the flow equations are as the following:
\begin{eqnarray}
&&\frac{da}{d\ell}=a,\cr&& \frac{dr}{d\ell}=2r+\frac{2\alpha}{\sqrt{1+r}}%
\cdot I_3(a), \cr &&\frac{d\alpha}{d\ell}=\epsilon\alpha-\frac{5}{2}\frac{%
\alpha^2} {(1+r)^\frac{3}{2}}\cdot I_3(a).
\end{eqnarray}

Then we have a new fixed point at $(r^\ast, \alpha^\ast, a^\ast)=(-\frac{2}{5%
}\epsilon,\frac{2}{5I_3(0)}\epsilon,0)$. Around this fixed point we define $%
r=r^\ast+\delta r$, $\alpha=\alpha^\ast+\delta\alpha$, $a=a^\ast+\delta a$
and have the linearized equations,%
\begin{eqnarray}
\frac{d}{d\ell}\left(%
\begin{array}{c}
\delta r \\
\delta\alpha \\
\delta a%
\end{array}%
\right)=\left(%
\begin{array}{ccc}
2-\frac{2}{5}\epsilon & 2 I_3(0) & \frac{4}{5 I_3(0)}\frac{\partial I_3(a)}{%
\partial a}|_{a=0} \\
0 & -\epsilon & 0 \\
0 & 0 & 1%
\end{array}%
\right)\left(%
\begin{array}{c}
\delta r \\
\delta\alpha \\
\delta a%
\end{array}%
\right).
\end{eqnarray}
The eigenvalues are $2-\frac{2}{5}\epsilon$, $-\epsilon$, and $1$. In three
dimensions the correlation length exponent is $\nu=\frac{1}{2-\frac{2}{5}%
\epsilon}=\frac{5}{9}$.

\end{widetext}


\begin{thebibliography}{99}
\bibitem{higher}
We also verified numerically that qualitatively our results are not changed by including these processes and higher bands.
\bibitem{2d}
Though it is not accurate, if one still applies the perturbation expansion to two-dimension case, it will predict a critical exponent $\nu=1/(2-2\epsilon/5)=5/7$ different from the mean-field value.
\end{thebibliography}
\end{document}